\documentclass[aps,prc,twocolumn,showpacs,nofootinbib]{revtex4-2}

\usepackage[dvips]{color}
\usepackage{ulem}
\usepackage{graphicx}
\usepackage[tbtags]{amsmath}
\usepackage{bm,amssymb}

\newcommand{\br}{{\bm{r}}}
\newcommand{\bp}{{\bm{p}}}

\newcommand{\sK}{\mathcal{K}}

\newcommand{\eps}{\varepsilon}

\newcommand{\nuc}[2]{{\mbox{$^{#2}${#1}}}}
\def\<{\langle}
\def\>{\rangle}
\renewcommand{\emph}[1]{\textit{#1}}

\begin{document}
\title{Octupole deformation of nuclei near the spherical closed-shell
configurations}
\author{Ken-ichiro Arita}
\email{arita@nitech.ac.jp}
\affiliation{Department of Physics, Nagoya Institute of Technology,
Nagoya 466-8555, Japan}
\accepted{June 12, 2023}

\begin{abstract}
The origin of octupole deformation for even-even nuclei near the
doubly-closed shell configurations are investigated by means of the
semiclassical periodic orbit theory.  In order to focus on the change
of shell structure due to deformation, a simple infinite-well
potential model is employed with octupole shape parameterized by
merging a sphere and a paraboloid.  Attention is paid to the
contributions of the degenerate families of periodic orbits (POs)
confined in the spherical portion of the potential, that are expected
to partially preserve the spherical shell effect up to considerably
large value of the octupole parameter.  The contribution of those POs
to the semiclassical trace formula plays an important role in bringing
about shell energy gain due to octupole deformation in the system with
a few particles added to spherical closed-shell configurations.
\end{abstract}
\pacs{21.60.-n, 36.40.-c, 03.65.Sq, 05.45.Mt}
\maketitle

\section{Introduction}

Atomic nuclei take various shapes with varying numbers of constituent
protons and neutrons, and the single-particle shell structures play
the essential role in their deformations and shape stabilities.  In
general, systems with particle numbers sufficiently far from the
spherical magic numbers will deform.  The majority of the ground-state
shapes are known to be quadrupole type, but some exotic shapes are found
depending on the combinations of proton and neutron numbers, and the
possible breaking of the reflection symmetry is one of the fundamental
problems in nuclear structure physics.  The ground-state octupole
deformations are observed only for a few nuclei, such as those
around the neutron-rich Ba region and Ra-Th region.  These regions are
located in the ``north-eastern'' neighbors of doubly magic nuclei on the
$(N,Z)$ plane of the nuclear chart, namely, they correspond to the
systems with a few particles added to spherical closed-shell
configurations \cite{Butler96}.  Possible static octupole shapes for
even-even nuclei have been systematically investigated with various
theoretical approaches such as microscopic-macroscopic
models \cite{Moller08}, the generator coordinate method \cite{Robledo11},
density functional theories \cite{Agbem16,Cao20}, and recently with the
Hartree-Fock-BCS model with three-dimensional Cartesian mesh
representation \cite{Ebata17}, which are consistent with the
experimental data and suggest promising regions of nuclei where
octupole deformation might be found.

\begin{figure}[tb]
\centering
\includegraphics[width=.8\linewidth]{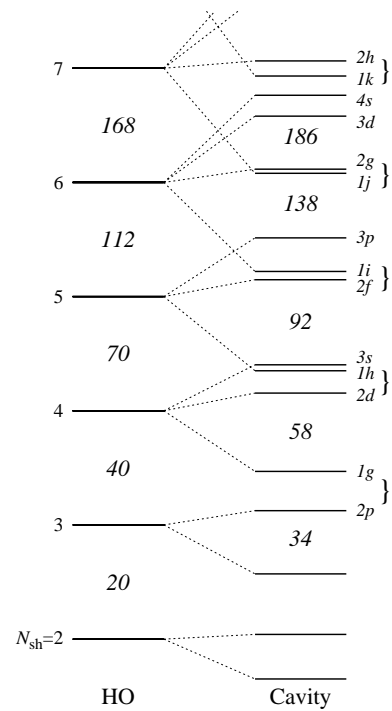}
\caption{\label{fig:spscomp}
Comparison of the single-particle spectra for spherical harmonic
oscillator (HO) potential model (left) and spherical infinite-well
(cavity) potential model (right).  Nearly degenerate $\varDelta l=3$
pairs of the single-particle levels in the cavity model are indicated
by braces.  The italic numbers marked on each energy gap represent the
number of levels below the gap, taking spin degeneracy into account.}
\end{figure}

As the origin of the ground-state octupole deformations for these
nuclei, the octupole correlation within the approximately degenerate
$\varDelta l=3$ pair of single-particle levels is considered to play a
significant role.  Such pairs of levels arise systematically
above the spherical
shell gaps for systems with sharp surface potential.
For example, $(2g_{7/2}, 1j_{15/2})$ orbitals above the $N(Z)=126$ gap
and $(2f_{5/2}, 1i_{13/2})$ orbitals above the $N(Z)=82$ gap are
approximately degenerate in the realistic nuclear mean field potential.

Figure~\ref{fig:spscomp} compares the single-particle spectra of the spherical
harmonic-oscillator (HO) potential model and the spherical
infinite-well potential (cavity) model, which have been referred to as
schematic models for light and heavy nuclei, respectively.
In the cavity model, one finds
pairs of $\varDelta l=3$ levels (enclosed by braces) above each shell
gap, for instance, $2g$ and $1j$ levels above the $N=138$ gap, $2f$
and $1i$ levels above the $N=92$ gap.
Thus, the cavity potential preserves
important features of the shell structure of the realistic nuclear mean field,
although the magic numbers are a little shifted from those of the
realistic ones due to the absence of spin-orbit coupling.

Since one has large octupole matrix elements between such $\varDelta
l=3$ levels, one of the levels is expected to go down rapidly
with increasing octupole deformation,
and the system
just above the closed-shell configuration which occupies this downward
level would prefer octupole shape\cite{Butler96}.
The behavior of those levels with respect to perturbations of the
octupole operators and their relation to the octupole deformation
energy have been examined in Refs.~\cite{HamMot91,Frisk94}.

On the other hand, from the view point of the shell correction method,
shell energy is governed by the gross shell structure\cite{Strut76}
and it is not obvious whether the origin of total shell
energy can be attributed to the behavior
of specific orbitals.  Moreover, there must be some
simple mechanism involved in the remarkable systematics in the
distribution of reflection asymmetry on the nuclear chart found in the
above numerical calculations.

In this paper, I analyze a simple cavity potential model to reveal
the essential mechanism of the nuclear octupole deformation.
As well as the behavior of $\varDelta l=3$ pairs
of single-particle
levels, I shall consider the effect of gross shell structure
from a semiclassical point of view; namely, I examine the role of the
classical periodic orbits (POs) in the semiclassical single-particle
level density.

The idea of this work was brought about by my recent works with
my colleagues, in
which we discussed the deformed shell effect of nuclei
through the fission path
\cite{Arita18A,Arita18B,Arita20}.
In the fission process,
a nucleus is elongated and a neck is formed which gradually separates the
system into two subsystems.  Such subsystems are called
\textit{prefragments}.  The prefragment
shell effect, associated with each of the subsystems, is expected to
come up after the neck formation\cite{Mosel71A,Mosel71B},
and it must be playing a significant
role in determining the fission path in the deformation space and the
resulting fragment mass distribution.
However, it is usually difficult to extract the prefragment effect
alone out of the total shell effect since most of the single-particle
wave functions are not localized in each of the prefragments.
To deal with this problem, we have proposed a simple idea using the
semiclassical periodic orbit theory (POT)\cite{Arita18B}.  In the
semiclassical trace formula, shell energy is expressed as the sum over
contributions of classical POs.  When the neck is formed, one has
families of POs confined in each of the prefragments, and their
contributions to the level density can be regarded as the prefragment
shell effect.  The POs in the spherical (but truncated) prefragment
make a strong shell effect similar to (but a little smaller than) that
for a full spherical potential.  Since the POs with the
same property have the same kind of contribution to the shell energy,
the prefragment PO should bring about considerable shell-energy gain
to the system when the size of the prefragment is same as that of the
spherical magic nucleus.  Such a condition for the sizes of prefragments
is favored by the nucleus in the fission process, and this provides
a simple and intuitive explanation for the mechanism of the asymmetric
fission in actinide nuclei.  Although the cavity model employed in the
above work is unrealistic, especially just before the scission point
for instance, the essential mechanism for the prefragment shell effect
will be applicable in more realistic situations.  In the realistic
density functional theory calculation, it has been shown
that the nucleon distributions in the prefragments for the fissioning
nucleus are very similar to those of isolated
nuclei\cite{Zhang16,Sadhukhan17}.  Then,
one expects the same mean field in a prefragment as that for an isolated
nucleus, and the semiclassical mechanism of the prefragment shell effect
associated with the classical POs localized in the prefragment seems to
be justified.  One can expect the same situation in nuclei just
above the spherical shell closures.

Thus, the main issue of this paper is to show that the above idea of
the prefragment shell effect can be also used in explaining the
systematics of the octupole deformation.  For this aim, I employ a
simple cavity potential model whose surface shape is made of a sphere
and a paraboloid joined together.  In the study of octupole deformation,
the surface shape is usually expanded in terms of spherical harmonics
$Y_{lm}$ for convenience\cite{FraPas96,Moller08}.  On the other hand,
the way of introducing reflection asymmetry in this work is based on
the physical
insight that the nucleus would have shell energy gain associated with
the spherical subsystem, in the same way as the strongly elongated
nuclei in the fission
processes.  The comparison of this parametrization with the conventional
one is made in the separate paper\cite{Arita23B}.

Apart from the above objective, I would also like to consider two other
problems using this model.  The first is to answer the question whether
the octupole deformation of the cavity boundary causes the parity
mixing of approximately degenerate $\varDelta l=3$ levels in the same way
as the perturbation of the potential by the octupole operator.
It is a nontrivial question which cannot be simply answered by the
ordinary method of perturbation.  The second is to confirm the
validity of the semiclassical trace formula for the truncated
spherical cavity which I have developed\cite{Arita18A}.

This paper is organized as follows.  In Sec.~\ref{sec:delta_l_3}, the
octupole cavity potential model employed in this work is defined, and
details of the shape parametrization are discussed.  Then, I
investigate the parity mixing of the $\varDelta l=3$ pair of
single-particle levels.  Next, in Sec.~\ref{sec:pot}, I consider the
evolution of gross shell structure with increasing octupole
deformation, and its role in explaining the systematics of the
octupole deformation is analyzed with
the use of the semiclassical POT.  Section~\ref{sec:summary} is devoted
to the summary and concluding remarks.

\section{Octupole correlation between the $\varDelta l=3$ pair of levels}
\label{sec:delta_l_3}

\begin{figure}
\centering
\includegraphics[width=.6\linewidth]{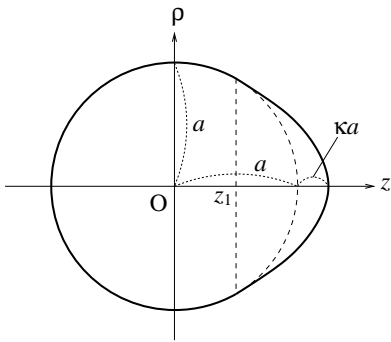}
\caption{\label{fig:shape}
Shape of the octupole surface defined by Eq.~(\ref{eq:shape}).  A
sphere and a paraboloid are smoothly joined at $z=z_1$.  The relative
size of the ``tip'' part $\kappa$ is considered as the octupole
parameter.}
\end{figure}

As illustrated in Fig.~\ref{fig:shape}, octupole deformation can be
induced by pinching one spot on the surface of the sphere.
Here, I shall use the term ``octupole deformation'' symbolically as
the shape with finite octupole moment.  In general, expansion of the
reflection-asymmetric nuclear surface shape into the spherical harmonics
can contain higher order multipole components ($Y_{lm}$ with $l>3$)
but the main reflection-asymmetric component must be $Y_{3m}$.
I shall parameterize the axially symmetric octupole shape by merging a
sphere and a paraboloid.  Then, the surface
$\rho=\rho_s(z)$ in the cylindrical coordinate $(\rho,\varphi,z)$
is expressed  as
\begin{gather}
\rho_s^2(z)=\left\{\begin{array}{l@{\qquad}l}
 a^2-z^2, & -a\leq z\leq z_1, \\
 2z_1((1+\kappa)a-z), & z_1\leq z\leq (1+\kappa)a,
                   \end{array}\right. \nonumber \\
z_1=\left(1+\kappa-\sqrt{(2+\kappa)\kappa}\right)a \label{eq:shape}
\end{gather}
where the sphere and the paraboloid are smoothly merged at $z=z_1$.
The thickness $\kappa(\geq 0)$ of the paraboloid ``tip'' relative to the
radius $a$ of the sphere part can be regarded as the octupole
parameter.  The parameter $a$ is determined so that the volume
conservation condition is satisfied.  Such shape parametrization is
initiated
to obtain a shell effect originated from the contribution of
classical PO families confined in the spherical subsystem.

\begin{figure}[tb]
\includegraphics[width=.8\linewidth]{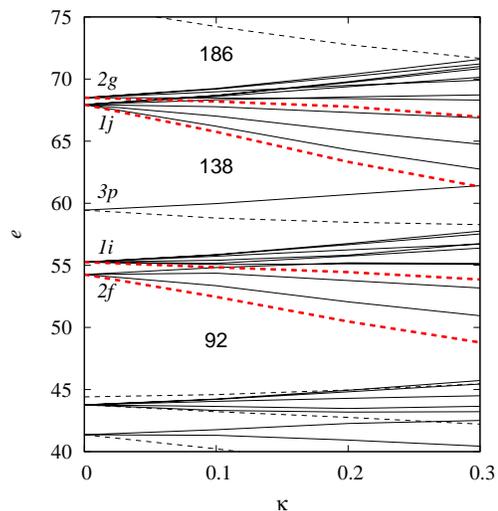}
\caption{\label{fig:ndiag}
Single-particle level diagram.  Broken and solid curves represent
$K=0$ and $K\geq 1$ levels, respectively.  Thick broken curves
represent the pairs of $\varDelta l=3$ levels with $K=0$ that are
approximately degenerate in the spherical limit $\kappa=0$.}
\end{figure}

Let us first look at the single-particle shell structure in the above
octupole-deformed infinite-well potential
\begin{equation}
V(\br)=\left\{\begin{array}{c@{\qquad}l}
 0 & [\rho(z) \leq \rho_s(z)] \\
 +\infty & [\rho(z) > \rho_s(z)]
              \end{array}\right.
\end{equation}
Figure~\ref{fig:ndiag}
shows the single-particle level diagram plotted against the
octupole parameter $\kappa$.
The eigenvalue problem for the Laplace equation with Dirichlet
boundary condition can be solved, e.g., by the method described
in Ref.~\cite{MukPal95}, which has been taken here.
At above each of the spherical gaps such as $N=92$ and $138$, one may
find the levels rapidly go down with increasing $\kappa$.
Let us examine the reasons of such behavior.
The most rapidly decreasing level above the $N=92$ (138) gap is the
$2f$ ($1j$) orbital with the magnetic quantum number $K=0$, and there are
$\varDelta l=3$ orbitals $1i$ ($2g$) just above it (see also Fig.~\ref{fig:spscomp}).
These pairs are indicated by the thick broken curves in
Fig.~\ref{fig:ndiag}.  Thus, the above behavior of the levels seems to
be related to the parity mixing of those levels due to the octupole
correlation.

Here let us review some basics on the breaking of reflection symmetry
and parity mixing.  Suppose that two levels $|1\>$ and $|2\>$ with
opposite parities are approximately degenerate in the symmetric limit:
\begin{gather}
H_0|1\>=(\eps-\delta)|1\>, \quad H_0|2\>=(\eps+\delta)|2\>
\quad (\delta>0), \nonumber \\
P|1\>=\sigma |1\>, \quad P|2\>=-\sigma|2\> \quad (\sigma=\pm 1),
\end{gather}
where $P$ is the parity (space inversion) operator and $H_0$ is the
reflection-symmetric Hamiltonian ($H_0P=PH_0$).  Consider the
parity-violating perturbation $\lambda V$ ($PV=-VP$) which satisfies
$\<1|V|1\>=\<2|V|2\>=0$ and $\<1|V|2\>=\<2|V|1\>=v>0$.  Then, the parity
mixing is described by the $2\times 2$ Hamiltonian matrix
\begin{equation}
H=H_0+\lambda V=\begin{pmatrix}\eps-\delta & \lambda v \\
  \lambda v & \eps+\delta\end{pmatrix}.
\end{equation}
The solutions of the eigenvalue equation $H|\psi_\pm\>=E_\pm|\psi_\pm\>$
are given by
\begin{gather}
E_\pm(\lambda)=\eps\pm\sqrt{\delta^2+(\lambda v)^2}, \\
|\psi_-(\lambda)\>=C\left(|1\>-\frac{\lambda v}{\sqrt{\delta^2
 +(\lambda v)^2}+\delta}|2\>\right), \nonumber \\
|\psi_+(\lambda)\>=C\left(|2\>+\frac{\lambda v}{\sqrt{\delta^2
 +(\lambda v)^2}+\delta}|1\>\right),
\end{gather}
where $C$ is the normalization constant.
The parity doublet $|1\>$ and $|2\>$ gradually mix and the energy
splitting grows with increasing
$\lambda$.  Finally, for $\lambda v\gg\delta$, complete mixing is
achieved and one has the parity partner
\begin{gather}
|\psi_\pm\>\simeq\frac{1}{\sqrt{2}}(|1\>\pm |2\>), \quad
P|\psi_\pm\>\simeq\sigma|\psi_\mp\>, \label{eq:parity}
\end{gather}
where the energy splitting is given approximately by
\begin{equation}
\varDelta E\sim 2\lambda v. \label{eq:parity_splitting}
\end{equation}

\begin{figure}
\includegraphics[width=.9\linewidth]{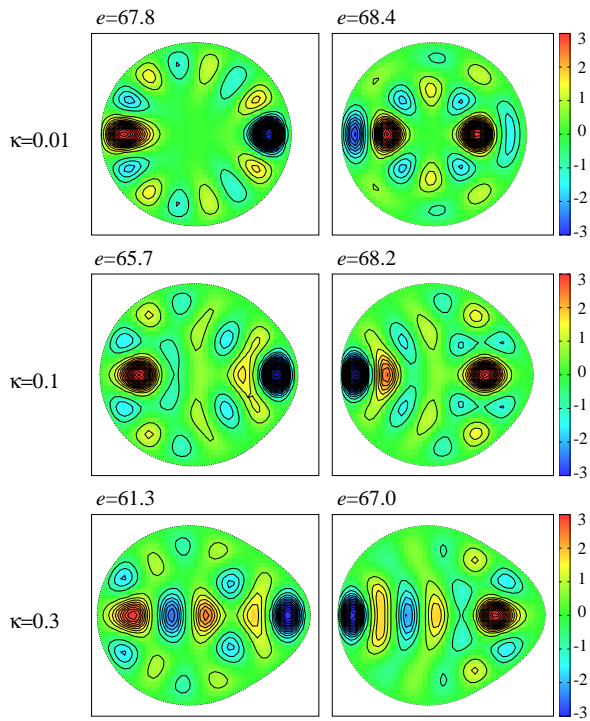}
\caption{\label{fig:wfun}
Evolution of the wave functions of the $\varDelta l=3$ pair of
single-particle levels just above the $N=138$ gap.  The contour plots
of the wave functions are shown.
The energy eigenvalue $e$ of each level is given at the upper
left of the figure.
The panels in the
left and right columns are for the lower and upper levels, originating
from the negative-parity $1j$ and positive-parity $2g$ orbits,
respectively, with increasing octupole parameter $\kappa$ from top to
bottom.}
\end{figure}

The question here is whether the changes in wave functions and
eigenvalue energies as described above also apply to the
cavity model against octupole deformation of the potential surface.
Figure~\ref{fig:wfun} displays the evolutions of the wave functions of a
pair of single-particle levels with increasing $\kappa$ for the $K=0$
states originating from the $\varDelta l=3$ orbital pair $1j$ and $2g$ just
above the spherical gap $N=138$.  Each panel shows the contour plot of
the wave function.  At $\kappa=0.01$, each wave
function is almost entirely occupied by the parity eigenstate.  With
increasing octupole deformation, a complete mixing seems to be
achieved already at $\kappa\approx 0.1$, where one of the wave functions
is quite similar to a space inversion of the other as shown in the second
equation of (\ref{eq:parity}).  The behavior of the energy splitting
with increasing $\kappa$ is also consistent with
Eq.~(\ref{eq:parity_splitting}) by assuming $\lambda\propto \kappa$.
The same properties also hold for the $K>0$ pairs of levels.  Thus, it
is confirmed that the parity mixing of nearly degenerate $\varDelta l=3$
levels explains the behavior of the single-particle shell structure
against the octupole deformation.  This behavior is expected to play a
certain role in enabling the system to achieve stable octupole
deformation.

\section{Gross shell structure in terms of classical periodic orbits}
\label{sec:pot}

The liquid drop model explains
an average property of nuclei, and the quantum fluctuation about it
is essentially given by the single-particle shell effect.
In a liquid drop picture, a nucleus is most stable in the spherical
shape, which minimizes the surface energy.  
The pronounced shell structure in the spherical potential is
advantageous for the closed-shell configurations, and conversely,
disadvantageous for the open-shell configurations.
The spherical shape becomes more unstable as the number of particles
deviates from any magic number corresponding to the closed-shell
configuration, and the system will deform when the shell
energy gain due to the deformation 
surpasses the increase of liquid-drop surface energy.

The nuclear ground-state deformations are considered
to be of the quadrupole type in most cases.
Another reason why the
quadrupole type deformation is most likely to occur is
the regularity of single-particle motion, which contributes to the
strong deformed shell effect.
In a potential with small quadrupole deformation, classical motion
of a single particle is mostly regular (stable).  However, the classical
motion rapidly becomes chaotic (unstable) with increasing octupole-type
deformation\cite{Heiss94}.  In general, quantum level repulsion
occurs in a classically chaotic system, which makes the shell effect
small compared to systems where the classical motion is regular.
For an exotic deformation to emerge, a considerably strong shell effect is
necessary which is usually associated with
dynamical symmetries, or resonances in another word, arising
locally in the system for
specific potential shapes\cite{Arita95,Sugita98,Arita16}.
A typical example is the so-called superdeformed state,
where the axis ratio is approximately 2:1.  It is understood in analogy
with the pronounced degeneracy of levels found
in a deformed oscillator potential with rational axis ratio.

In analyzing the origin of such gross shell effect, semiclassical POT
provides us with a powerful tool\cite{Gutz71,BaBlo3,BBBook,Arita12,Arita16}.
In general, distribution of single-particle energy eigenvalues shows a
regular oscillating pattern, but its origin cannot be explained within
the framework of pure quantum mechanics.  To describe the above
oscillation, Balian and Bloch considered a semiclassical
approximation and derived an outstanding
formula which expresses the quantum level density
\begin{equation}
g(e)=\sum_{i}\delta(e-e_i)
\end{equation}
as the sum over contributions from the classical POs\cite{BaBlo3}.
The formula they have obtained
is specific to the infinite-well (cavity) potential
systems, although it is applicable to any dimension and shape.  Independently
of them, Gutzwiller derived the same type of the formula
from a different semiclassical approach\cite{Gutz71}.  His
formula, known as the Gutzwiller trace formula, can be applied to
Hamiltonian systems with more generic potentials, but is
limited to the case where all classical motions are unstable
and the system has no continuous symmetries such as
rotational symmetries.
In the opposite extreme with respect to the stability of the
classical motions,
the trace formulas for completely integrable (multiply periodic) systems
are derived by Berry and Tabor based on the torus quantization condition
of Einstein, Brillouin, and Keller (EBK)~\cite{BerTab76}.  The extension of
the Gutzwiller trace formula to systems with continuous symmetries
was made, e.g., in Refs.~\cite{Strut76,Creagh91}.  The formulas
applicable to the stable orbits that encounter bifurcations, for which
Gutzwiller's formula breaks down, have been derived by the uniform
approximations\cite{Schom97} and the improved stationary-phase
approximation\cite{MagISP1,MagISP2}.  The general version of the trace
formula, incorporating all the above, might be expressed as
\begin{gather}
g(e)=g_0(e)+\delta g(e), \\
\delta g(e)\simeq \sum_{\rm PO}A_{\rm PO}(e)\cos\left(\tfrac{1}{\hslash}
 S_{\rm PO}(e)-\tfrac{\pi}{2}\mu_{\rm PO}\right), \label{eq:trace_g}
\end{gather}
where $g_0$ represents the average level density, equivalent to the
(extended) Thomas-Fermi approximation\cite{Jenn74,RSBook,BBBook}, and
the oscillating component $\delta g$ is expressed as the sum over
the contribution of classical POs.  $S_{\rm PO}=\oint_{\rm PO}\bp\cdot
d\br$ represents the action integral along the PO, $\mu_{\rm PO}$ is
the Maslov index related to the geometrical character of PO, and the
amplitude $A_{\rm PO}$ is fully determined by the classical properties (such
as degeneracy, period, and stability) of the orbit.  Since the action
integral is generally a monotonically increasing function of energy
$e$, each contribution of PO in the right-hand side of
Eq.~(\ref{eq:trace_g}) gives a regularly oscillating function of $e$.
The orbit with shorter period $T_{\rm PO}=dS_{\rm PO}/de$ gives the
gross structure of the level density and the longer orbits contribute
to the finer structures.  In order to investigate the gross shell
structure, it is sufficient to consider the contributions of only a
few shortest POs.  If the single-particle Hamiltonian has continuous
symmetries, each PO generally forms a continuous family of several
parameters.  Such a family is called a degenerate orbit and the number
of continuous parameters $\sK_{\rm PO}$ for the family is called the
degeneracy.  Note that the orbits with higher degeneracies make a more
significant contribution to the level density.  Speaking in the
context of semiclassical $\hslash$ expansion, the amplitude factor
$A_{\rm PO}$ is of the order $\hslash^{-\sK_{\rm PO}/2}$.

Looking at the level diagram in Fig.~\ref{fig:ndiag}, one will find an
approximately degenerate cluster of levels below each spherical shell
gap, preserving strong shell effects up to fairly large values of the
octupole parameter $\kappa$ .  As I show in the following, this strong
shell effect under octupole deformation is related to the local
symmetry of the system, namely, the presence of the partially
spherically symmetric subsystem.  In the smooth potential models,
dynamical symmetries play the same role.
If the system has such special local
symmetry or dynamical symmetry under the exotic shape, a strong deformed
shell effect is expected and the importance of such shape degree of
freedom might come into competition with that of the quadrupole type.

Among the classical POs in the cavity model under consideration, there
are degenerate family of orbits localized in the sphere part of the
potential.
\begin{figure}
\centering
\includegraphics[viewport=25 5 505 225,width=\linewidth]{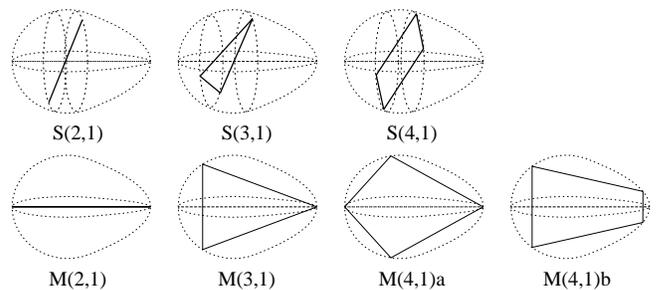}
\caption{\label{fig:po_wqs}
Some short classical POs in the octupole cavity whose surface is given
by Eq.~(\ref{eq:shape}) with the octupole parameter $\kappa=0.2$.  The
lower panels represent the meridian-plane orbits M($v,w$), and the
upper panels represent the regular polygon orbit families S($v,w$)
confined in the sphere part of the potential.  The indices $v$ and $w$
represent the number of vertices and the winding number, respectively.}
\end{figure}
Figure~\ref{fig:po_wqs} displays some short simple POs.  The upper
panels show the diameter ($\sK=2$) and regular polygon
($\sK=3$) families of orbits localized in the spherical part, and
the lower panels show the isolated ($\sK=0$) linear symmetry-axis orbit
and the meridian-plane orbit families ($\sK=1$).  There are also
three-dimensional (non-planar) orbits that form $\sK=1$ families, but they
are longer than the above ones and contribute only to the finer shell
structures.

One can see the contribution of these orbits to the semiclassical level
density using the Fourier transformation technique.  Through the
classical motion of the particle in the cavity potential, the magnitude
of the momentum $p$ is kept constant,
and the action integral along the orbit is simply
given by the product of $p$ and the geometric length $L_{\rm PO}$
of the orbit.
Thus, the level density in the wave-number variable $k$ ($p=\hslash k$)
is expressed as
\begin{align}
g(k)&=g(e)\frac{de}{dk} \nonumber \\
&=g_0(k)+\sum_{\rm PO}a_{\rm PO}(k)
\cos(kL_{\rm PO}-\tfrac{\pi}{2}\mu_{\rm PO}). \label{eq:trace_gk}
\end{align}
The simple $k$ dependence of the above phase factor enables us
to estimate the contribution of each orbit by the Fourier
transformation of level density.  Let us consider the Fourier transform
defined by
\begin{gather}
F(L)=\sqrt{\frac{\pi}{2}}\frac{1}{k_c}
\int g(k)e^{ikL}e^{-(k/k_c)^2/2}dk. \label{eq:fourier}
\end{gather}
In this definition, a Gaussian cutoff factor is
incorporated into the integrand in order to exclude the high energy
part ($k\gg k_c$) of the level density which is numerically
inaccessible.  The calculation of the Fourier transform of the exact
quantum level density is straightforward if one has the quantum energy
spectrum $\{e_j=(\hslash k_j)^2/2m\}$.  Inserting
$g(k)=\sum_j\delta(k-k_j)$ into Eq.~(\ref{eq:fourier}), one has
\begin{equation}
F^{\rm (qm)}(L)=\sqrt{\frac{\pi}{2}}\frac{1}{k_c}\sum_j
e^{ik_jL}e^{-(k_j/k_c)^2/2}. \label{eq:fourier_qm}
\end{equation}
On the other hand, by inserting the
semiclassical expression (\ref{eq:trace_gk}) into (\ref{eq:fourier}),
one has
\begin{equation}
F^{\rm (cl)}(L)=F_0(L)+\sum_{\rm PO}a_{\rm PO}e^{-i\pi\mu_{\rm PO}/2}
e^{-\{k_c(L-L_{\rm PO})\}^2/2} \label{eq:scf}
\end{equation} 
which is a function exhibiting peaks at the lengths of the POs,
$L=L_{\rm PO}$, with heights proportional to the amplitude $a_{\rm
PO}$.  In deriving Eq.~(\ref{eq:scf}), $k$ dependence of the amplitude
$a_{\rm PO}$ is ignored for simplicity.  Taking into account the
correct $k$ dependence, one has another expression where the Gaussian
is replaced by a different but similar single-peaked function (see
Fig.~11 of Ref.~\cite{Arita18A}).

In this way, one can extract information on the contribution of
classical POs by the Fourier transform of the quantum level density.
The summation in Eq.~(\ref{eq:fourier_qm}) can be truncated at certain
$k_{\rm max}$ if one takes $k_c$ sufficiently smaller than $k_{\rm
max}$.  $k_c$ determines the resolution $\varDelta L$ of the orbit
length by the uncertainty relation $\varDelta L=1/k_c$.  Sufficiently
large $k_c$ is required for a good resolution of the orbit length, and
I took $k_cR_0=20$ ($R_0$ being the radius of the potential in the
spherical limit) and $k_{\rm max}=\tfrac32 k_c$ in the present
calculation.

\begin{figure}
\centering
\includegraphics[width=\linewidth]{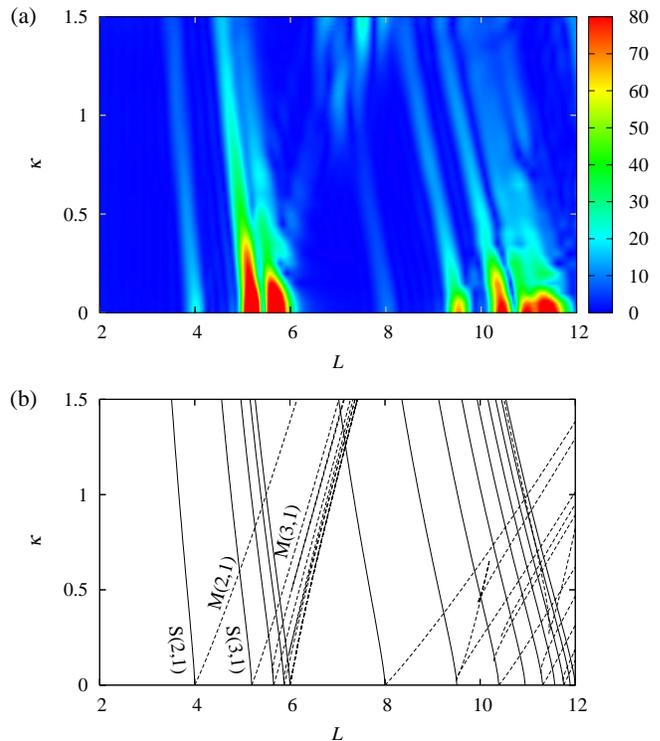}
\caption{\label{fig:fmap2}
In the upper panel (a), modulus of the Fourier transform of the quantum
level density $|F^{\rm (qm)}(L;\kappa)|$ [see Eq.~(\ref{eq:fourier_qm})]
is shown as a function of $L$ and
$\kappa$.  In the lower panel (b), lengths of the classical POs are
plotted as functions of $\kappa$.  Solid lines represent the lengths
of the regular polygon orbits confined in the sphere part, and the
broken lines are for the meridian-plane orbits.}
\end{figure}

The upper panel of Fig.~\ref{fig:fmap2} displays the modulus of
quantum-mechanical Fourier transform $|F^{\rm (qm)}(L;\kappa)|$ as a
function of the length variable
$L$ and the octupole parameter $\kappa$.  In the lower panel, 
the length of the classical POs are plotted as functions of
octupole parameter.  Solid curves represent the lengths of the regular
polygon POs confined
in the sphere part of the potential, and broken lines represent those
of the meridian-plane orbits.
By comparing these two panels, it can be seen
that the Fourier amplitude has strong peaks mainly along the orbit
families confined in the sphere part.
Particularly, the peak corresponding to the triangular orbit S(3,1) is
outstanding.  Thus, one can expect that the gross
shell effect is given mostly by the contribution of this triangular
family.

The effect of the shell structure on deformation should be
estimated by the shell energy, rather than the level density.
Using Eq.~(\ref{eq:trace_g}), one obtains the trace formula for
shell energy as\cite{Strut76,BBBook}
\begin{align}
&\delta E(N)=\int^{e_F}(e-e_F)\delta g(e)de \nonumber \\
&~ \simeq\sum_{\rm PO}\frac{\hslash^2}{T_{\rm PO}^2}
A_{\rm PO}(e_F)\cos\left(k_FL_{\rm PO}
-\tfrac{\pi}{2}\mu_{\rm PO}\right), \label{eq:trace_sce}
\end{align}
where $e_F=(\hslash k_F)^2/2M$ is the Fermi energy satisfying
\begin{equation}
N=\int^{e_F}g(e)de.
\end{equation}
The additional factor $T_{\rm PO}^{-2}$ in
Eq.~(\ref{eq:trace_sce}) suppresses the contributions of longer
orbits, and accordingly one has only to consider a few
shortest POs with higher degeneracies.

For the cavity model under consideration, the contribution of the PO
family confined in the sphere part can be directly evaluated by the
trace formula for a truncated spherical cavity which has been
derived for the study of the nascent-fragment (prefragment) shell
effect in nuclear fission processes\cite{Arita18A}.
\begin{figure}
\centering
\includegraphics[width=\linewidth]{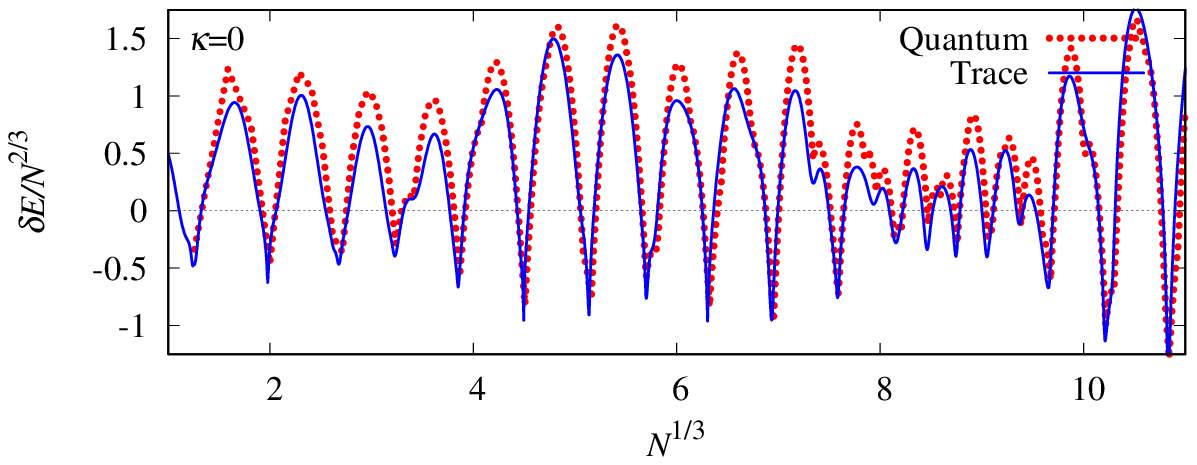}\\
\includegraphics[width=\linewidth]{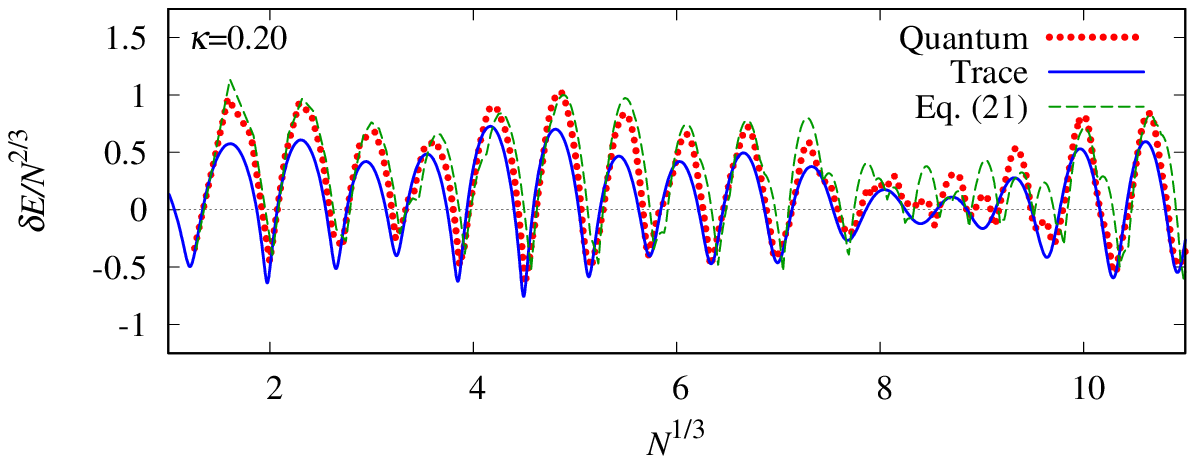}\\
\includegraphics[width=\linewidth]{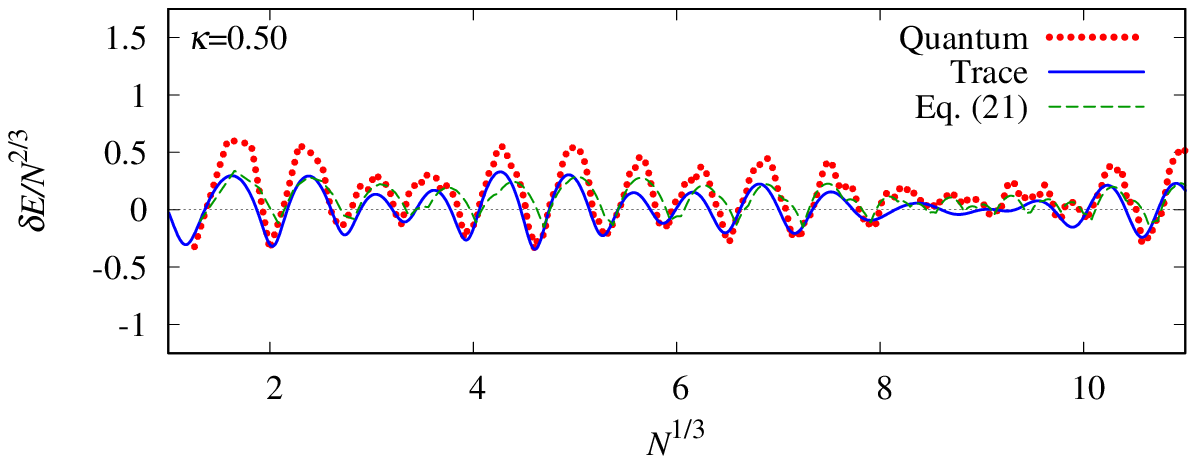}
\caption{\label{fig:trace_wqs}
Shell energy $\delta E(N)$ multiplied by $N^{-2/3}$ plotted
against $N^{1/3}$, for the octupole parameter values
$\kappa=0$, 0.2, and 0.5.  Dotted curve represent the quantum result,
and the solid curve represent the semiclassical trace formula
taking into account the contribution of some short regular polygon
families confined in the sphere part of the cavity.}
\end{figure}
Figure~\ref{fig:trace_wqs} shows the results of shell energies
(\ref{eq:trace_sce}) for the octupole parameter values $\kappa=0$, 0.2, and
0.5.  For these relatively small octupole deformations,
quantum results are nicely reproduced by the contribution of POs
confined in the sphere part of the potential.  One finds that the
oscillating pattern in case of the spherical shape
survives well in the octupole deformed system.

Keeping in mind that the shell effect is essentially determined by the
POs confined in the sphere part of the potential,
let us consider the condition for the system to take the
octupole shape by focusing attention on the PO contribution.
Because of the saturation property, volume $V$ surrounded by the
potential surface is proportional to the particle number $N$.
According to the Weyl's asymptotic formula\cite{BaBlo1}, the
leading term of the average level density is given by\footnote[1]{%
The unit $\hslash^2/2M=1$ is used in Ref.~\cite{BaBlo1}.}
\begin{equation}
g_0(e)\simeq \frac{2M}{\hslash^2}\frac{Vk}{4\pi^2}, \label{eq:18}
\end{equation}
with the volume
\begin{equation}
V=\frac{4\pi}{3}R_0^3=\frac{4\pi}{3}Nr_0^3,
\end{equation}
where $R_0=N^{1/3}r_0$ is the nuclear
radius in the spherical limit.  From the relation between Fermi wave
number $k_F$ and particle number $N$, one obtains
\begin{gather*}
N\simeq \int_0^{\hslash^2k_F^2/2M}g_0(e)de=\frac{Vk_F^3}{6\pi^2}
 \simeq \frac{2N(k_Fr_0)^3}{9\pi}, \\
k_F\simeq \left(\frac{9\pi}{2}\right)^{1/3}r_0^{-1}.
\end{gather*}
Thus, the value of the Fermi wave number $k_F$ is approximately
independent of the particle number $N$.  

In the trace formula (\ref{eq:trace_sce}), let us introduce the
reduction factor $w_{\rm PO}$ of the PO family amplitude due to the
truncation, and assume that the Maslov indices are unchanged by the
truncation\footnote[2]{To be precise, one has the small shifts of the Maslov
indices due to the contribution of the marginal orbits which
correspond to the higher-order quantum corrections\cite{Arita18A}.}
\begin{equation}
A_{\rm PO}(e_F)=w_{\rm PO}A_{\rm PO}^{(0)}(e_F), \quad
\mu_{\rm PO}\simeq \mu_{\rm PO}^{(0)},
\end{equation}
where $A_{\rm PO}^{(0)}$ and $\mu_{\rm PO}^{(0)}$ represent the
amplitude and Maslov indices for the PO family in the
spherical cavity without truncation.
Inserting them into Eq.~(\ref{eq:trace_sce}) and
replacing $w_{\rm PO}$ with $w_{31}$
of the dominant triangular orbit S(3,1), one has
\begin{align}
\delta E(N) &\simeq w_{31}\sum_{\rm PO}\frac{\hslash^2}{T_{\rm PO}^2}
 A_{\rm PO}^{(0)}\cos(k_FL_{\rm PO}-\tfrac{\pi}{2}\mu_{\rm PO})
 \nonumber \\
 &=w_{31}\delta E^{(0)}(N^{(0)}(e_F)), \label{eq:qmcomp}
\end{align}
where $\delta E^{(0)}$ is the shell energy of the spherical cavity
without truncation.
Since the number of the constituent particles is
proportional to the volume surrounded by the potential surface under the fixed
Fermi energy, one has
\begin{equation}
\frac{N^{(0)}(e_F)}{N}=\frac{V_{\rm sph}}{V}\equiv f(\kappa),
\label{eq:fkappa}
\end{equation}
where $V$ and $V_{\rm sph}$ are volumes of the total system and that
of the sphere composing the octupole surface (\ref{eq:shape}), $V_{\rm
sph}=4\pi a^3/3$.  $f(\kappa)$ is a monotonically decreasing function of
$\kappa$ as easily presumed from Fig.~\ref{fig:shape}.
As displayed in Fig.~\ref{fig:trace_wqs}, the expression of
Eq.~(\ref{eq:qmcomp}) explains the main feature of the shell structure
quite well.

\begin{figure}
\includegraphics[width=\linewidth]{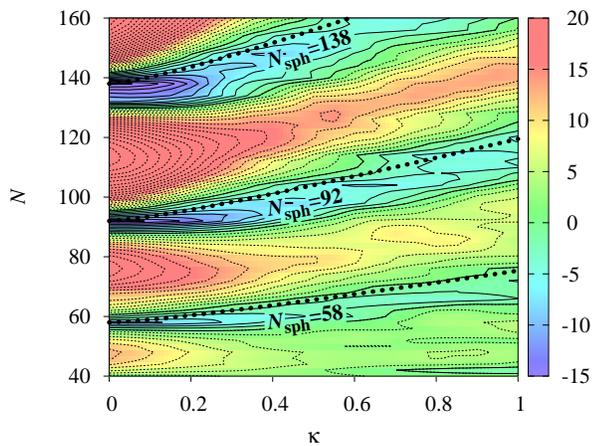}
\caption{\label{fig:sce}
Contour plot of the shell energies as functions of octupole parameter
$\kappa$ and particle number $N$.  Solid and broken contour curves
represent negative and positive shell energies, respectively.
Thick curves show where the radius of the sphere part of the potential
is equal to the radius of the spherical magic nuclei
with particle numbers $N_{\rm sph}=58$, $92$ and $138$.}
\end{figure}

According to the rough but meaningful estimation discussed above,
the shell energy described
by the contribution of PO in Eq.~(\ref{eq:trace_sce}) is
essentially governed by the lengths $L_{\rm PO}$ of the orbits
in the sphere part of the potential.
The contribution of those POs will then
minimize the shell energy when the radius of the sphere part is
identical to the radius of spherical magic nucleus.
Figure~\ref{fig:sce} shows the contour plot of the shell energy in the
$(\kappa,N)$ plane.  The curves (\ref{eq:fkappa}) for $N^{(0)}$ at
some spherical magic numbers $N_{\rm sph}$ are also drawn in the
figure.  One will find that those curves successfully explain the shell
energy valleys.

\section{Summary}
\label{sec:summary}

Octupole deformation of nuclei above the spherical magic
configurations are investigated by the simple cavity potential model,
where the potential surface is parameterized by merging a sphere and a
paraboloid.  The semiclassical trace formula for the truncated
spherical cavity is successfully applied to our model and gives us
a clear understanding of the properties of shell structure.
The contribution of degenerate orbit family confined in the
spherical subsystem
brings about a strong
shell effect similar to those in the spherical shape, and it plays a
significant role in stabilizing the octupole shape.
This mechanism nicely explains the systematics of the octupole
deformations on the nuclear chart.

This result is also related to the recent works on the role of the
octupole shape degree of freedom in fission
fragments\cite{Scamps18,Scamps19}.  In the fission process,
prefragments take octupole shapes near the scission point, and the
octupole shell effect controls the size of the fragment.  Since
particle numbers a little above the spherical magic number prefer
octupole deformation, it explains why the mass number of heavier
fragment is concentrated around 140, a little larger than that of
doubly magic \nuc{Sn}{132}.

The current shape parametrization can be generalized to
spheroid+paraboloid, which enables us to investigate the ground state
shapes of nuclei, taking account of the quadrupole and octupole shape
degrees of freedom.  Results of the systematic analysis with such an
extension will be discussed in a separate paper\cite{Arita23B}.

There have been various approaches to examine the ground-state
octupole deformation over the nuclear chart, and in most of those
analyses, axially symmetric type of octupole
deformation was the main consideration.  In this work, I have
also limited myself to the axially symmetric case.
However, it should be mentioned that nonaxial
octupole shape degrees of freedom and the role of $\varDelta l=3$
pair of levels in it were analyzed, and a pronounced bunching of levels
was found in the case of $Y_{32}$ deformation, which has tetrahedral
symmetry\cite{HamMot91,Frisk94}.
The theoretical search of tetrahedral nuclei has been extensively carried
out with the realistic mean field model\cite{Dudek02,Dudek18}.
Recently, all four types of octupole shapes and the role of the
point-group symmetries were examined in Pb and superheavy
regions\cite{Yang22A,Yang22B}.

In the present work, the axially symmetric octupole deformation for
the nuclei just
above the spherical shell closures is shown to be related to the
dynamical symmetry, which can be taken as a partial
survival of the spherical symmetry for special combination of
quadrupole and octupole deformations.
On the other hand, a strong tetrahedral shell effect is expected
by the bifurcation of PO on the way from
spherical to larger tetrahedral deformation\cite{AriMuk14}.
It is an interesting subject to investigate the systematics of
nonaxial octupole deformations over the nuclear chart and its
semiclassical origin, which is left for future work.

\acknowledgments

I would like to thank Prof. Kenichi Matsuyanagi for his
helpful comments and discussions.  Part of the numerical calculations
in this work were carried out at the Yukawa Institute Computer Facility.

\bibliographystyle{apsrev4-2}
\bibliography{refs}

\begin{thebibliography}{42}%
\makeatletter
\providecommand \@ifxundefined [1]{%
 \@ifx{#1\undefined}
}%
\providecommand \@ifnum [1]{%
 \ifnum #1\expandafter \@firstoftwo
 \else \expandafter \@secondoftwo
 \fi
}%
\providecommand \@ifx [1]{%
 \ifx #1\expandafter \@firstoftwo
 \else \expandafter \@secondoftwo
 \fi
}%
\providecommand \natexlab [1]{#1}%
\providecommand \enquote  [1]{``#1''}%
\providecommand \bibnamefont  [1]{#1}%
\providecommand \bibfnamefont [1]{#1}%
\providecommand \citenamefont [1]{#1}%
\providecommand \href@noop [0]{\@secondoftwo}%
\providecommand \href [0]{\begingroup \@sanitize@url \@href}%
\providecommand \@href[1]{\@@startlink{#1}\@@href}%
\providecommand \@@href[1]{\endgroup#1\@@endlink}%
\providecommand \@sanitize@url [0]{\catcode `\\12\catcode `\$12\catcode
  `\&12\catcode `\#12\catcode `\^12\catcode `\_12\catcode `\%12\relax}%
\providecommand \@@startlink[1]{}%
\providecommand \@@endlink[0]{}%
\providecommand \url  [0]{\begingroup\@sanitize@url \@url }%
\providecommand \@url [1]{\endgroup\@href {#1}{\urlprefix }}%
\providecommand \urlprefix  [0]{URL }%
\providecommand \Eprint [0]{\href }%
\providecommand \doibase [0]{https://doi.org/}%
\providecommand \selectlanguage [0]{\@gobble}%
\providecommand \bibinfo  [0]{\@secondoftwo}%
\providecommand \bibfield  [0]{\@secondoftwo}%
\providecommand \translation [1]{[#1]}%
\providecommand \BibitemOpen [0]{}%
\providecommand \bibitemStop [0]{}%
\providecommand \bibitemNoStop [0]{.\EOS\space}%
\providecommand \EOS [0]{\spacefactor3000\relax}%
\providecommand \BibitemShut  [1]{\csname bibitem#1\endcsname}%
\let\auto@bib@innerbib\@empty
\bibitem [{\citenamefont {Butler}\ and\ \citenamefont
  {Nazarewicz}(1996)}]{Butler96}%
  \BibitemOpen
  \bibfield  {author} {\bibinfo {author} {\bibfnamefont {P.~A.}\ \bibnamefont
  {Butler}}\ and\ \bibinfo {author} {\bibfnamefont {W.}~\bibnamefont
  {Nazarewicz}},\ }\href@noop {} {\bibfield  {journal} {\bibinfo  {journal}
  {Rev. Mod. Phys.}\ }\textbf {\bibinfo {volume} {68}},\ \bibinfo {pages} {349}
  (\bibinfo {year} {1996})}\BibitemShut {NoStop}%
\bibitem [{\citenamefont {M\"{o}ller}\ \emph {et~al.}(2008)\citenamefont
  {M\"{o}ller}, \citenamefont {Bengtsson}, \citenamefont {Carlsson},
  \citenamefont {Olivius}, \citenamefont {Ichikawa}, \citenamefont {Sagawa},\
  and\ \citenamefont {Iwamoto}}]{Moller08}%
  \BibitemOpen
  \bibfield  {author} {\bibinfo {author} {\bibfnamefont {P.}~\bibnamefont
  {M\"{o}ller}}, \bibinfo {author} {\bibfnamefont {R.}~\bibnamefont
  {Bengtsson}}, \bibinfo {author} {\bibfnamefont {B.}~\bibnamefont {Carlsson}},
  \bibinfo {author} {\bibfnamefont {P.}~\bibnamefont {Olivius}}, \bibinfo
  {author} {\bibfnamefont {T.}~\bibnamefont {Ichikawa}}, \bibinfo {author}
  {\bibfnamefont {H.}~\bibnamefont {Sagawa}},\ and\ \bibinfo {author}
  {\bibfnamefont {A.}~\bibnamefont {Iwamoto}},\ }\href@noop {} {\bibfield
  {journal} {\bibinfo  {journal} {Atomic Data and Nuclear Data Tables}\
  }\textbf {\bibinfo {volume} {94}},\ \bibinfo {pages} {758} (\bibinfo {year}
  {2008})}\BibitemShut {NoStop}%
\bibitem [{\citenamefont {Robledo}\ and\ \citenamefont
  {Bertsch}(2011)}]{Robledo11}%
  \BibitemOpen
  \bibfield  {author} {\bibinfo {author} {\bibfnamefont {L.~M.}\ \bibnamefont
  {Robledo}}\ and\ \bibinfo {author} {\bibfnamefont {G.~F.}\ \bibnamefont
  {Bertsch}},\ }\href@noop {} {\bibfield  {journal} {\bibinfo  {journal} {Phys.
  Rev. C}\ }\textbf {\bibinfo {volume} {84}},\ \bibinfo {pages} {054302}
  (\bibinfo {year} {2011})}\BibitemShut {NoStop}%
\bibitem [{\citenamefont {Agbemava}\ \emph {et~al.}(2016)\citenamefont
  {Agbemava}, \citenamefont {Afanasjev},\ and\ \citenamefont {Ring}}]{Agbem16}%
  \BibitemOpen
  \bibfield  {author} {\bibinfo {author} {\bibfnamefont {S.~E.}\ \bibnamefont
  {Agbemava}}, \bibinfo {author} {\bibfnamefont {A.~V.}\ \bibnamefont
  {Afanasjev}},\ and\ \bibinfo {author} {\bibfnamefont {P.}~\bibnamefont
  {Ring}},\ }\href@noop {} {\bibfield  {journal} {\bibinfo  {journal} {Phys.
  Rev. C}\ }\textbf {\bibinfo {volume} {93}},\ \bibinfo {pages} {044304}
  (\bibinfo {year} {2016})}\BibitemShut {NoStop}%
\bibitem [{\citenamefont {Cao}\ \emph {et~al.}(2020)\citenamefont {Cao},
  \citenamefont {Agbemava}, \citenamefont {Afanasjev}, \citenamefont
  {Nazarewicz},\ and\ \citenamefont {Olsen}}]{Cao20}%
  \BibitemOpen
  \bibfield  {author} {\bibinfo {author} {\bibfnamefont {Y.}~\bibnamefont
  {Cao}}, \bibinfo {author} {\bibfnamefont {S.~E.}\ \bibnamefont {Agbemava}},
  \bibinfo {author} {\bibfnamefont {A.~V.}\ \bibnamefont {Afanasjev}}, \bibinfo
  {author} {\bibfnamefont {W.}~\bibnamefont {Nazarewicz}},\ and\ \bibinfo
  {author} {\bibfnamefont {E.}~\bibnamefont {Olsen}},\ }\href@noop {}
  {\bibfield  {journal} {\bibinfo  {journal} {Phys. Rev. C}\ }\textbf {\bibinfo
  {volume} {102}},\ \bibinfo {pages} {024311} (\bibinfo {year}
  {2020})}\BibitemShut {NoStop}%
\bibitem [{\citenamefont {Ebata}\ and\ \citenamefont
  {Nakatsukasa}(2017)}]{Ebata17}%
  \BibitemOpen
  \bibfield  {author} {\bibinfo {author} {\bibfnamefont {S.}~\bibnamefont
  {Ebata}}\ and\ \bibinfo {author} {\bibfnamefont {T.}~\bibnamefont
  {Nakatsukasa}},\ }\href@noop {} {\bibfield  {journal} {\bibinfo  {journal}
  {Phys. Scr.}\ }\textbf {\bibinfo {volume} {92}},\ \bibinfo {pages} {064006}
  (\bibinfo {year} {2017})}\BibitemShut {NoStop}%
\bibitem [{\citenamefont {Hamamoto}\ \emph {et~al.}(1991)\citenamefont
  {Hamamoto}, \citenamefont {Mottelson}, \citenamefont {Xie},\ and\
  \citenamefont {Zhang}}]{HamMot91}%
  \BibitemOpen
  \bibfield  {author} {\bibinfo {author} {\bibfnamefont {I.}~\bibnamefont
  {Hamamoto}}, \bibinfo {author} {\bibfnamefont {B.~R.}\ \bibnamefont
  {Mottelson}}, \bibinfo {author} {\bibfnamefont {H.}~\bibnamefont {Xie}},\
  and\ \bibinfo {author} {\bibfnamefont {X.~Z.}\ \bibnamefont {Zhang}},\
  }\href@noop {} {\bibfield  {journal} {\bibinfo  {journal} {Z. Phys. D}\
  }\textbf {\bibinfo {volume} {21}},\ \bibinfo {pages} {163} (\bibinfo {year}
  {1991})}\BibitemShut {NoStop}%
\bibitem [{\citenamefont {Frisk}\ \emph {et~al.}(1994)\citenamefont {Frisk},
  \citenamefont {Hamamoto},\ and\ \citenamefont {May}}]{Frisk94}%
  \BibitemOpen
  \bibfield  {author} {\bibinfo {author} {\bibfnamefont {F.}~\bibnamefont
  {Frisk}}, \bibinfo {author} {\bibfnamefont {I.}~\bibnamefont {Hamamoto}},\
  and\ \bibinfo {author} {\bibfnamefont {F.~R.}\ \bibnamefont {May}},\
  }\href@noop {} {\bibfield  {journal} {\bibinfo  {journal} {Phys. Scr.}\
  }\textbf {\bibinfo {volume} {50}},\ \bibinfo {pages} {628} (\bibinfo {year}
  {1994})}\BibitemShut {NoStop}%
\bibitem [{\citenamefont {Strutinsky}\ and\ \citenamefont
  {Magner}(1976)}]{Strut76}%
  \BibitemOpen
  \bibfield  {author} {\bibinfo {author} {\bibfnamefont {V.~M.}\ \bibnamefont
  {Strutinsky}}\ and\ \bibinfo {author} {\bibfnamefont {A.~G.}\ \bibnamefont
  {Magner}},\ }\href@noop {} {\bibfield  {journal} {\bibinfo  {journal} {Sov.
  J. Part. Nucl.}\ }\textbf {\bibinfo {volume} {7}},\ \bibinfo {pages} {138}
  (\bibinfo {year} {1976})}\BibitemShut {NoStop}%
\bibitem [{\citenamefont {Arita}(2018)}]{Arita18A}%
  \BibitemOpen
  \bibfield  {author} {\bibinfo {author} {\bibfnamefont {K.}~\bibnamefont
  {Arita}},\ }\href@noop {} {\bibfield  {journal} {\bibinfo  {journal} {Phys.
  Rev. C}\ }\textbf {\bibinfo {volume} {98}},\ \bibinfo {pages} {064310}
  (\bibinfo {year} {2018})}\BibitemShut {NoStop}%
\bibitem [{\citenamefont {Arita}\ \emph {et~al.}(2018)\citenamefont {Arita},
  \citenamefont {Ichikawa},\ and\ \citenamefont {Matsuyanagi}}]{Arita18B}%
  \BibitemOpen
  \bibfield  {author} {\bibinfo {author} {\bibfnamefont {K.}~\bibnamefont
  {Arita}}, \bibinfo {author} {\bibfnamefont {T.}~\bibnamefont {Ichikawa}},\
  and\ \bibinfo {author} {\bibfnamefont {K.}~\bibnamefont {Matsuyanagi}},\
  }\href@noop {} {\bibfield  {journal} {\bibinfo  {journal} {Phys. Rev. C}\
  }\textbf {\bibinfo {volume} {98}},\ \bibinfo {pages} {064311} (\bibinfo
  {year} {2018})}\BibitemShut {NoStop}%
\bibitem [{\citenamefont {Arita}\ \emph {et~al.}(2020)\citenamefont {Arita},
  \citenamefont {Ichikawa},\ and\ \citenamefont {Matsuyanagi}}]{Arita20}%
  \BibitemOpen
  \bibfield  {author} {\bibinfo {author} {\bibfnamefont {K.}~\bibnamefont
  {Arita}}, \bibinfo {author} {\bibfnamefont {T.}~\bibnamefont {Ichikawa}},\
  and\ \bibinfo {author} {\bibfnamefont {K.}~\bibnamefont {Matsuyanagi}},\
  }\href@noop {} {\bibfield  {journal} {\bibinfo  {journal} {Phys. Scr.}\
  }\textbf {\bibinfo {volume} {95}},\ \bibinfo {pages} {024003} (\bibinfo
  {year} {2020})}\BibitemShut {NoStop}%
\bibitem [{\citenamefont {Mosel}\ and\ \citenamefont
  {Schmitt}(1971{\natexlab{a}})}]{Mosel71A}%
  \BibitemOpen
  \bibfield  {author} {\bibinfo {author} {\bibfnamefont {U.}~\bibnamefont
  {Mosel}}\ and\ \bibinfo {author} {\bibfnamefont {H.~W.}\ \bibnamefont
  {Schmitt}},\ }\href@noop {} {\bibfield  {journal} {\bibinfo  {journal} {Nucl.
  Phys. A}\ }\textbf {\bibinfo {volume} {165}},\ \bibinfo {pages} {73}
  (\bibinfo {year} {1971}{\natexlab{a}})}\BibitemShut {NoStop}%
\bibitem [{\citenamefont {Mosel}\ and\ \citenamefont
  {Schmitt}(1971{\natexlab{b}})}]{Mosel71B}%
  \BibitemOpen
  \bibfield  {author} {\bibinfo {author} {\bibfnamefont {U.}~\bibnamefont
  {Mosel}}\ and\ \bibinfo {author} {\bibfnamefont {H.~W.}\ \bibnamefont
  {Schmitt}},\ }\href@noop {} {\bibfield  {journal} {\bibinfo  {journal} {Phys.
  Rev. C}\ }\textbf {\bibinfo {volume} {4}},\ \bibinfo {pages} {2185} (\bibinfo
  {year} {1971}{\natexlab{b}})}\BibitemShut {NoStop}%
\bibitem [{\citenamefont {Zhang}\ \emph {et~al.}(2016)\citenamefont {Zhang},
  \citenamefont {Schuetrumpf},\ and\ \citenamefont {Nazarewicz}}]{Zhang16}%
  \BibitemOpen
  \bibfield  {author} {\bibinfo {author} {\bibfnamefont {C.~L.}\ \bibnamefont
  {Zhang}}, \bibinfo {author} {\bibfnamefont {B.}~\bibnamefont {Schuetrumpf}},\
  and\ \bibinfo {author} {\bibfnamefont {W.}~\bibnamefont {Nazarewicz}},\
  }\href@noop {} {\bibfield  {journal} {\bibinfo  {journal} {Phys. Rev. C}\
  }\textbf {\bibinfo {volume} {94}},\ \bibinfo {pages} {064323} (\bibinfo
  {year} {2016})}\BibitemShut {NoStop}%
\bibitem [{\citenamefont {Sadhukhan}\ \emph {et~al.}(2017)\citenamefont
  {Sadhukhan}, \citenamefont {l.~Zhang}, \citenamefont {Nazarewicz},\ and\
  \citenamefont {Schunck}}]{Sadhukhan17}%
  \BibitemOpen
  \bibfield  {author} {\bibinfo {author} {\bibfnamefont {J.}~\bibnamefont
  {Sadhukhan}}, \bibinfo {author} {\bibfnamefont {C.}~\bibnamefont {l.~Zhang}},
  \bibinfo {author} {\bibfnamefont {W.}~\bibnamefont {Nazarewicz}},\ and\
  \bibinfo {author} {\bibfnamefont {N.}~\bibnamefont {Schunck}},\ }\href@noop
  {} {\bibfield  {journal} {\bibinfo  {journal} {Phys. Rev. C}\ }\textbf
  {\bibinfo {volume} {96}},\ \bibinfo {pages} {061301(R)} (\bibinfo {year}
  {2017})}\BibitemShut {NoStop}%
\bibitem [{\citenamefont {Frauendorf}\ and\ \citenamefont
  {Pashkevich}(1996)}]{FraPas96}%
  \BibitemOpen
  \bibfield  {author} {\bibinfo {author} {\bibfnamefont {S.}~\bibnamefont
  {Frauendorf}}\ and\ \bibinfo {author} {\bibfnamefont {V.~V.}\ \bibnamefont
  {Pashkevich}},\ }\href@noop {} {\bibfield  {journal} {\bibinfo  {journal}
  {Ann. Phys. (Berlin)}\ }\textbf {\bibinfo {volume} {508}},\ \bibinfo {pages}
  {34} (\bibinfo {year} {1996})}\BibitemShut {NoStop}%
\bibitem [{\citenamefont {Arita}(2023)}]{Arita23B}%
  \BibitemOpen
  \bibfield  {author} {\bibinfo {author} {\bibfnamefont {K.}~\bibnamefont
  {Arita}},\ }\href@noop {} {\bibfield  {journal} {\bibinfo  {journal}
  {Preprint arXiv:2304.00655}\ } (\bibinfo {year} {2023})}\BibitemShut
  {NoStop}%
\bibitem [{\citenamefont {Mukhopadhyay}\ and\ \citenamefont
  {Pal}(1995)}]{MukPal95}%
  \BibitemOpen
  \bibfield  {author} {\bibinfo {author} {\bibfnamefont {T.}~\bibnamefont
  {Mukhopadhyay}}\ and\ \bibinfo {author} {\bibfnamefont {S.}~\bibnamefont
  {Pal}},\ }\href@noop {} {\bibfield  {journal} {\bibinfo  {journal} {Nucl.
  Phys. A}\ }\textbf {\bibinfo {volume} {592}},\ \bibinfo {pages} {291}
  (\bibinfo {year} {1995})}\BibitemShut {NoStop}%
\bibitem [{\citenamefont {Heiss}\ \emph {et~al.}(1994)\citenamefont {Heiss},
  \citenamefont {Nazmitdinov},\ and\ \citenamefont {Radu}}]{Heiss94}%
  \BibitemOpen
  \bibfield  {author} {\bibinfo {author} {\bibfnamefont {W.~D.}\ \bibnamefont
  {Heiss}}, \bibinfo {author} {\bibfnamefont {R.~G.}\ \bibnamefont
  {Nazmitdinov}},\ and\ \bibinfo {author} {\bibfnamefont {S.}~\bibnamefont
  {Radu}},\ }\href@noop {} {\bibfield  {journal} {\bibinfo  {journal} {Phys.
  Rev. Lett}\ }\textbf {\bibinfo {volume} {72}},\ \bibinfo {pages} {2351}
  (\bibinfo {year} {1994})}\BibitemShut {NoStop}%
\bibitem [{\citenamefont {Arita}\ and\ \citenamefont
  {Matsuyanagi}(1995)}]{Arita95}%
  \BibitemOpen
  \bibfield  {author} {\bibinfo {author} {\bibfnamefont {K.}~\bibnamefont
  {Arita}}\ and\ \bibinfo {author} {\bibfnamefont {K.}~\bibnamefont
  {Matsuyanagi}},\ }\href@noop {} {\bibfield  {journal} {\bibinfo  {journal}
  {Nucl. Phys. A}\ }\textbf {\bibinfo {volume} {592}},\ \bibinfo {pages} {9}
  (\bibinfo {year} {1995})}\BibitemShut {NoStop}%
\bibitem [{\citenamefont {Sugita}\ \emph {et~al.}(1998)\citenamefont {Sugita},
  \citenamefont {Arita},\ and\ \citenamefont {Matsuyanagi}}]{Sugita98}%
  \BibitemOpen
  \bibfield  {author} {\bibinfo {author} {\bibfnamefont {A.}~\bibnamefont
  {Sugita}}, \bibinfo {author} {\bibfnamefont {K.}~\bibnamefont {Arita}},\ and\
  \bibinfo {author} {\bibfnamefont {K.}~\bibnamefont {Matsuyanagi}},\
  }\href@noop {} {\bibfield  {journal} {\bibinfo  {journal} {Prog. Theor.
  Phys.}\ }\textbf {\bibinfo {volume} {100}},\ \bibinfo {pages} {597} (\bibinfo
  {year} {1998})}\BibitemShut {NoStop}%
\bibitem [{\citenamefont {Arita}(2016)}]{Arita16}%
  \BibitemOpen
  \bibfield  {author} {\bibinfo {author} {\bibfnamefont {K.}~\bibnamefont
  {Arita}},\ }\href@noop {} {\bibfield  {journal} {\bibinfo  {journal} {Phys.
  Scr.}\ }\textbf {\bibinfo {volume} {91}},\ \bibinfo {pages} {063002}
  (\bibinfo {year} {2016})}\BibitemShut {NoStop}%
\bibitem [{\citenamefont {Gutzwiller}(1971)}]{Gutz71}%
  \BibitemOpen
  \bibfield  {author} {\bibinfo {author} {\bibfnamefont {M.~C.}\ \bibnamefont
  {Gutzwiller}},\ }\href@noop {} {\bibfield  {journal} {\bibinfo  {journal} {J.
  Math. Phys.}\ }\textbf {\bibinfo {volume} {12}},\ \bibinfo {pages} {343}
  (\bibinfo {year} {1971})}\BibitemShut {NoStop}%
\bibitem [{\citenamefont {Balian}\ and\ \citenamefont {Bloch}(1972)}]{BaBlo3}%
  \BibitemOpen
  \bibfield  {author} {\bibinfo {author} {\bibfnamefont {R.}~\bibnamefont
  {Balian}}\ and\ \bibinfo {author} {\bibfnamefont {C.}~\bibnamefont {Bloch}},\
  }\href@noop {} {\bibfield  {journal} {\bibinfo  {journal} {Ann. Phys. (NY)}\
  }\textbf {\bibinfo {volume} {69}},\ \bibinfo {pages} {76} (\bibinfo {year}
  {1972})}\BibitemShut {NoStop}%
\bibitem [{\citenamefont {Brack}\ and\ \citenamefont {Bhaduri}(2003)}]{BBBook}%
  \BibitemOpen
  \bibfield  {author} {\bibinfo {author} {\bibfnamefont {M.}~\bibnamefont
  {Brack}}\ and\ \bibinfo {author} {\bibfnamefont {R.~K.}\ \bibnamefont
  {Bhaduri}},\ }\href@noop {} {\emph {\bibinfo {title} {Semiclassical
  Physics}}}\ (\bibinfo  {publisher} {Westview Press, Boulder},\ \bibinfo
  {year} {2003})\BibitemShut {NoStop}%
\bibitem [{\citenamefont {Arita}(2012)}]{Arita12}%
  \BibitemOpen
  \bibfield  {author} {\bibinfo {author} {\bibfnamefont {K.}~\bibnamefont
  {Arita}},\ }\href@noop {} {\bibfield  {journal} {\bibinfo  {journal} {Phys.
  Rev. C}\ }\textbf {\bibinfo {volume} {86}},\ \bibinfo {pages} {034317}
  (\bibinfo {year} {2012})}\BibitemShut {NoStop}%
\bibitem [{\citenamefont {Berry}\ and\ \citenamefont {Tabor}(1976)}]{BerTab76}%
  \BibitemOpen
  \bibfield  {author} {\bibinfo {author} {\bibfnamefont {M.~V.}\ \bibnamefont
  {Berry}}\ and\ \bibinfo {author} {\bibfnamefont {M.}~\bibnamefont {Tabor}},\
  }\href@noop {} {\bibfield  {journal} {\bibinfo  {journal} {Proc. R. Soc.
  Lond. A}\ }\textbf {\bibinfo {volume} {349}},\ \bibinfo {pages} {101}
  (\bibinfo {year} {1976})}\BibitemShut {NoStop}%
\bibitem [{\citenamefont {Creagh}\ and\ \citenamefont
  {Littlejohn}(1991)}]{Creagh91}%
  \BibitemOpen
  \bibfield  {author} {\bibinfo {author} {\bibfnamefont {S.~C.}\ \bibnamefont
  {Creagh}}\ and\ \bibinfo {author} {\bibfnamefont {R.~G.}\ \bibnamefont
  {Littlejohn}},\ }\href@noop {} {\bibfield  {journal} {\bibinfo  {journal}
  {Phys. Rev. A}\ }\textbf {\bibinfo {volume} {44}},\ \bibinfo {pages} {836}
  (\bibinfo {year} {1991})}\BibitemShut {NoStop}%
\bibitem [{\citenamefont {Schomerus}\ and\ \citenamefont
  {Sieber}(1997)}]{Schom97}%
  \BibitemOpen
  \bibfield  {author} {\bibinfo {author} {\bibfnamefont {H.}~\bibnamefont
  {Schomerus}}\ and\ \bibinfo {author} {\bibfnamefont {M.}~\bibnamefont
  {Sieber}},\ }\href@noop {} {\bibfield  {journal} {\bibinfo  {journal} {J.
  Phys. A: Math. Gen.}\ }\textbf {\bibinfo {volume} {30}},\ \bibinfo {pages}
  {4537} (\bibinfo {year} {1997})}\BibitemShut {NoStop}%
\bibitem [{\citenamefont {Magner}\ \emph {et~al.}(1999)\citenamefont {Magner},
  \citenamefont {Fedotkin}, \citenamefont {Arita}, \citenamefont {Misu},
  \citenamefont {Matsuyanagi}, \citenamefont {Schachner},\ and\ \citenamefont
  {Brack}}]{MagISP1}%
  \BibitemOpen
  \bibfield  {author} {\bibinfo {author} {\bibfnamefont {A.~G.}\ \bibnamefont
  {Magner}}, \bibinfo {author} {\bibfnamefont {S.~N.}\ \bibnamefont
  {Fedotkin}}, \bibinfo {author} {\bibfnamefont {K.}~\bibnamefont {Arita}},
  \bibinfo {author} {\bibfnamefont {T.}~\bibnamefont {Misu}}, \bibinfo {author}
  {\bibfnamefont {K.}~\bibnamefont {Matsuyanagi}}, \bibinfo {author}
  {\bibfnamefont {T.}~\bibnamefont {Schachner}},\ and\ \bibinfo {author}
  {\bibfnamefont {M.}~\bibnamefont {Brack}},\ }\href@noop {} {\bibfield
  {journal} {\bibinfo  {journal} {Prog. Theor. Phys.}\ }\textbf {\bibinfo
  {volume} {102}},\ \bibinfo {pages} {551} (\bibinfo {year}
  {1999})}\BibitemShut {NoStop}%
\bibitem [{\citenamefont {Magner}\ and\ \citenamefont {Arita}(2017)}]{MagISP2}%
  \BibitemOpen
  \bibfield  {author} {\bibinfo {author} {\bibfnamefont {A.~G.}\ \bibnamefont
  {Magner}}\ and\ \bibinfo {author} {\bibfnamefont {K.}~\bibnamefont {Arita}},\
  }\href@noop {} {\bibfield  {journal} {\bibinfo  {journal} {Phys. Rev. E}\
  }\textbf {\bibinfo {volume} {96}},\ \bibinfo {pages} {042206} (\bibinfo
  {year} {2017})}\BibitemShut {NoStop}%
\bibitem [{\citenamefont {Jennings}(1974)}]{Jenn74}%
  \BibitemOpen
  \bibfield  {author} {\bibinfo {author} {\bibfnamefont {B.~K.}\ \bibnamefont
  {Jennings}},\ }\href@noop {} {\bibfield  {journal} {\bibinfo  {journal} {Ann.
  Phys. (NY)}\ ,\ \bibinfo {pages} {1}} (\bibinfo {year} {1974})}\BibitemShut
  {NoStop}%
\bibitem [{\citenamefont {Ring}\ and\ \citenamefont {Schuck}(1980)}]{RSBook}%
  \BibitemOpen
  \bibfield  {author} {\bibinfo {author} {\bibfnamefont {P.}~\bibnamefont
  {Ring}}\ and\ \bibinfo {author} {\bibfnamefont {P.}~\bibnamefont {Schuck}},\
  }\href@noop {} {\emph {\bibinfo {title} {The Nuclear Many-Body Problems}}}\
  (\bibinfo  {publisher} {Springer, New York},\ \bibinfo {year}
  {1980})\BibitemShut {NoStop}%
\bibitem [{\citenamefont {Balian}\ and\ \citenamefont {Bloch}(1970)}]{BaBlo1}%
  \BibitemOpen
  \bibfield  {author} {\bibinfo {author} {\bibfnamefont {R.}~\bibnamefont
  {Balian}}\ and\ \bibinfo {author} {\bibfnamefont {C.}~\bibnamefont {Bloch}},\
  }\href@noop {} {\bibfield  {journal} {\bibinfo  {journal} {Ann. Phys. (NY)}\
  }\textbf {\bibinfo {volume} {60}},\ \bibinfo {pages} {401} (\bibinfo {year}
  {1970})}\BibitemShut {NoStop}%
\bibitem [{\citenamefont {Scamps}\ and\ \citenamefont
  {Simenel}(2018)}]{Scamps18}%
  \BibitemOpen
  \bibfield  {author} {\bibinfo {author} {\bibfnamefont {G.}~\bibnamefont
  {Scamps}}\ and\ \bibinfo {author} {\bibfnamefont {C.}~\bibnamefont
  {Simenel}},\ }\href@noop {} {\bibfield  {journal} {\bibinfo  {journal}
  {Nature (London)}\ }\textbf {\bibinfo {volume} {564}},\ \bibinfo {pages}
  {382} (\bibinfo {year} {2018})}\BibitemShut {NoStop}%
\bibitem [{\citenamefont {Scamps}\ and\ \citenamefont
  {Simenel}(2019)}]{Scamps19}%
  \BibitemOpen
  \bibfield  {author} {\bibinfo {author} {\bibfnamefont {G.}~\bibnamefont
  {Scamps}}\ and\ \bibinfo {author} {\bibfnamefont {C.}~\bibnamefont
  {Simenel}},\ }\href@noop {} {\bibfield  {journal} {\bibinfo  {journal} {Phys.
  Rev. C}\ }\textbf {\bibinfo {volume} {100}},\ \bibinfo {pages} {041602(R)}
  (\bibinfo {year} {2019})}\BibitemShut {NoStop}%
\bibitem [{\citenamefont {Dudek}\ \emph {et~al.}(2002)\citenamefont {Dudek},
  \citenamefont {Go\'{z}d\'{z}}, \citenamefont {Schunck},\ and\ \citenamefont
  {Mi\'{s}kiewicz}}]{Dudek02}%
  \BibitemOpen
  \bibfield  {author} {\bibinfo {author} {\bibfnamefont {J.}~\bibnamefont
  {Dudek}}, \bibinfo {author} {\bibfnamefont {A.}~\bibnamefont
  {Go\'{z}d\'{z}}}, \bibinfo {author} {\bibfnamefont {N.}~\bibnamefont
  {Schunck}},\ and\ \bibinfo {author} {\bibfnamefont {M.}~\bibnamefont
  {Mi\'{s}kiewicz}},\ }\href@noop {} {\bibfield  {journal} {\bibinfo  {journal}
  {Phys. Rev. Lett.}\ }\textbf {\bibinfo {volume} {88}},\ \bibinfo {pages}
  {252502} (\bibinfo {year} {2002})}\BibitemShut {NoStop}%
\bibitem [{\citenamefont {Dudek}\ \emph {et~al.}(2018)\citenamefont {Dudek},
  \citenamefont {Curien}, \citenamefont {Dedes}, \citenamefont {k.~Mazurek},
  \citenamefont {Tagami}, \citenamefont {Shimizu},\ and\ \citenamefont
  {Bhattacharjee}}]{Dudek18}%
  \BibitemOpen
  \bibfield  {author} {\bibinfo {author} {\bibfnamefont {J.}~\bibnamefont
  {Dudek}}, \bibinfo {author} {\bibfnamefont {D.}~\bibnamefont {Curien}},
  \bibinfo {author} {\bibfnamefont {I.}~\bibnamefont {Dedes}}, \bibinfo
  {author} {\bibnamefont {k.~Mazurek}}, \bibinfo {author} {\bibfnamefont
  {S.}~\bibnamefont {Tagami}}, \bibinfo {author} {\bibfnamefont {Y.~R.}\
  \bibnamefont {Shimizu}},\ and\ \bibinfo {author} {\bibfnamefont
  {T.}~\bibnamefont {Bhattacharjee}},\ }\href@noop {} {\bibfield  {journal}
  {\bibinfo  {journal} {Phys. Rev. C}\ }\textbf {\bibinfo {volume} {97}},\
  \bibinfo {pages} {021302(R)} (\bibinfo {year} {2018})}\BibitemShut {NoStop}%
\bibitem [{\citenamefont {Yang}\ \emph
  {et~al.}(2022{\natexlab{a}})\citenamefont {Yang}, \citenamefont {Dudek},
  \citenamefont {Dedes}, \citenamefont {Baran}, \citenamefont {Curien},
  \citenamefont {Gaamouci}, \citenamefont {G{\'o}{\'z}d{\'z}}, \citenamefont
  {P\c{e}drak}, \citenamefont {Rouvel}, \citenamefont {Wang},\ and\
  \citenamefont {Burkat}}]{Yang22A}%
  \BibitemOpen
  \bibfield  {author} {\bibinfo {author} {\bibfnamefont {J.}~\bibnamefont
  {Yang}}, \bibinfo {author} {\bibfnamefont {J.}~\bibnamefont {Dudek}},
  \bibinfo {author} {\bibfnamefont {I.}~\bibnamefont {Dedes}}, \bibinfo
  {author} {\bibfnamefont {A.}~\bibnamefont {Baran}}, \bibinfo {author}
  {\bibfnamefont {D.}~\bibnamefont {Curien}}, \bibinfo {author} {\bibfnamefont
  {A.}~\bibnamefont {Gaamouci}}, \bibinfo {author} {\bibfnamefont
  {A.}~\bibnamefont {G{\'o}{\'z}d{\'z}}}, \bibinfo {author} {\bibfnamefont
  {A.}~\bibnamefont {P\c{e}drak}}, \bibinfo {author} {\bibfnamefont
  {D.}~\bibnamefont {Rouvel}}, \bibinfo {author} {\bibfnamefont {H.~L.}\
  \bibnamefont {Wang}},\ and\ \bibinfo {author} {\bibfnamefont
  {J.}~\bibnamefont {Burkat}},\ }\href@noop {} {\bibfield  {journal} {\bibinfo
  {journal} {Phys. Rev. C}\ }\textbf {\bibinfo {volume} {105}},\ \bibinfo
  {pages} {034348} (\bibinfo {year} {2022}{\natexlab{a}})}\BibitemShut
  {NoStop}%
\bibitem [{\citenamefont {Yang}\ \emph
  {et~al.}(2022{\natexlab{b}})\citenamefont {Yang}, \citenamefont {Dudek},
  \citenamefont {Dedes}, \citenamefont {Baran}, \citenamefont {Curien},
  \citenamefont {Gaamouci}, \citenamefont {G{\'o}{\'z}d{\'z}}, \citenamefont
  {P\c{e}drak}, \citenamefont {Rouvel},\ and\ \citenamefont {Wang}}]{Yang22B}%
  \BibitemOpen
  \bibfield  {author} {\bibinfo {author} {\bibfnamefont {J.}~\bibnamefont
  {Yang}}, \bibinfo {author} {\bibfnamefont {J.}~\bibnamefont {Dudek}},
  \bibinfo {author} {\bibfnamefont {I.}~\bibnamefont {Dedes}}, \bibinfo
  {author} {\bibfnamefont {A.}~\bibnamefont {Baran}}, \bibinfo {author}
  {\bibfnamefont {D.}~\bibnamefont {Curien}}, \bibinfo {author} {\bibfnamefont
  {A.}~\bibnamefont {Gaamouci}}, \bibinfo {author} {\bibfnamefont
  {A.}~\bibnamefont {G{\'o}{\'z}d{\'z}}}, \bibinfo {author} {\bibfnamefont
  {A.}~\bibnamefont {P\c{e}drak}}, \bibinfo {author} {\bibfnamefont
  {D.}~\bibnamefont {Rouvel}},\ and\ \bibinfo {author} {\bibfnamefont {H.~L.}\
  \bibnamefont {Wang}},\ }\href@noop {} {\bibfield  {journal} {\bibinfo
  {journal} {Phys. Rev. C}\ }\textbf {\bibinfo {volume} {106}},\ \bibinfo
  {pages} {054314} (\bibinfo {year} {2022}{\natexlab{b}})}\BibitemShut
  {NoStop}%
\bibitem [{\citenamefont {Arita}\ and\ \citenamefont
  {Mukumoto}(2014)}]{AriMuk14}%
  \BibitemOpen
  \bibfield  {author} {\bibinfo {author} {\bibfnamefont {K.}~\bibnamefont
  {Arita}}\ and\ \bibinfo {author} {\bibfnamefont {Y.}~\bibnamefont
  {Mukumoto}},\ }\href@noop {} {\bibfield  {journal} {\bibinfo  {journal}
  {Phys. Rev. C}\ }\textbf {\bibinfo {volume} {89}},\ \bibinfo {pages} {054308}
  (\bibinfo {year} {2014})}\BibitemShut {NoStop}%
\end{thebibliography}%
\end{document}